\documentstyle[11pt,paspconf,epsf]{article}
\def\edcomment#1{\iffalse\marginpar{\raggedright\sl#1\/}\else\relax\fi}
\reversemarginpar

\begin{document}
\title{Weak Lensing Mass of Nearby Clusters of Galaxies}
\author{Michael Joffre\altaffilmark{1,2}}
\author{Phillipe Fischer\altaffilmark{3}, Joshua
Frieman\altaffilmark{1,2}, David Johnston\altaffilmark{1,2}, Tim
McKay\altaffilmark{3}, Joseph Mohr\altaffilmark{2}, Bob
Nichol\altaffilmark{4}, Erin Sheldon\altaffilmark{3}, Chris
Cantaloupo\altaffilmark{4}, Greg Griffin\altaffilmark{4}, Jeff Peterson\altaffilmark{4}, Kathy Romer\altaffilmark{4} }

\altaffiltext{1}{ NASA/Fermilab Astrophysics Center, Fermi National Accelerator
Laboratory, Batavia, IL}
\altaffiltext{2}{Dept. of Astronomy and Astrophysics, University of Chicago, Chicago, IL}
\altaffiltext{3}{Dept. of Physics, University of Michigan, Ann Arbor, MI}
\altaffiltext{4}{Physics Department, Carnegie Mellon University, Pittsburgh, PA}

\begin{abstract}
We describe first results of a project to create weak lensing mass maps for a
complete, X-ray luminosity-limited sample of 19 nearby ($z<0.1$) southern galaxy
clusters scheduled for Sunyaev--Zel'dovich (SZ) observations by the Viper Telescope
at the South Pole.  We have collected data on 1/3 of the sample and present
motivation for the project as well as projected mass maps of two clusters.
\end{abstract}


Detection of weak gravitational lensing shear towards clusters of galaxies is now
routine (Clowe 1998, Kaiser 1998).  This tangential stretching of background galaxy images with
respect to the foreground mass distribution can be extracted to give two-dimensional
mass maps of galaxy clusters and yield determination of their total masses.  

Detecting this very weak tidal shearing of the light from background galaxies
requires averaging over many such galaxies to overcome the noise introduced by their
intrinsic ellipticity distribution.  If the rms ellipticity of these background galaxies
is $ e$ (typically .3), then the required number of galaxies $N$ to obtain a
signal to noise greater than one in the presence of a shear $s$ (typically $\sim  .05$ in
a cluster) is $N \ge (e/s)^2$ (Gould 1995).  To obtain these galaxy counts in the past
with standard size cameras, it was necessary to observe distant clusters with small
angular size to very faint magnitudes.   However, this precludes the study of nearby galaxy
clusters which have been studied extensively by other methods.  For nearby clusters,
it is possible to obtain the necessary galaxy counts by imaging a much wider, though
shallower, region of the sky.  

The possibility of covering large areas of the sky and thus observing nearby clusters
($ z < 0.1$) has become feasible with the advent of mosaic CCD cameras.  There are
several advantages of utilizing weak lensing to investigate nearby clusters over
distant clusters: 

-- Lensing strength of nearby clusters is very insensitive to the redshift of the
source objects.  This is not true of high redshift clusters, causing an uncertainty
in their mass estimate when the source redshifts are unknown.

--  Angular size and brightness of nearby background galaxies are larger, giving a higher
signal to noise for shape measurements than high-redshift galaxies.

-- There is a great deal of ancillary data on nearby clusters, such as optical
redshifts, X-ray data, and soon, high-resolution SZ data.  We gain most of our
knowledge of cluster properties from such data which is less available for high
redshift clusters.  

With these advantages in mind, we have begun a survey of 19 southern galaxy clusters
with $z <0.1$.  This X-ray luminosity-limited sample will be targeted by the Viper telescope 
for SZ observations (Romer 1999).  We have imaged 6 of the sample with the CTIO 4m telescope using
the BTC.  For each cluster, we have taken images in several filters over greater than
a $41'\times41'$ area to a depth of 25th magnitude in r.
To date, we have corrected two images for systematic errors and reconstructed a mass
map for each cluster (Figure 1). Abell 3667 is at $z=0.0530$ with a ROSAT X-ray
luminosity of $8.76\times10^{44}$ erg/s (Ebeling 1996).  Abell 3266 is at $z=0.0545$ with an X-ray
luminosity of $6.15\times 10^{44}$ ergs/s.  Such analysis of the entire sample should
present a consistent picture of mass concentrations in low redshift clusters.  It will also allow us to combine weak lensing information with other ancillary data such as SZ
and X-ray measurements.  Such combinations will enable far more detailed
studies of binding mass, baryon fractions, and morphologies of nearby clusters.
\begin{figure}
\plottwo{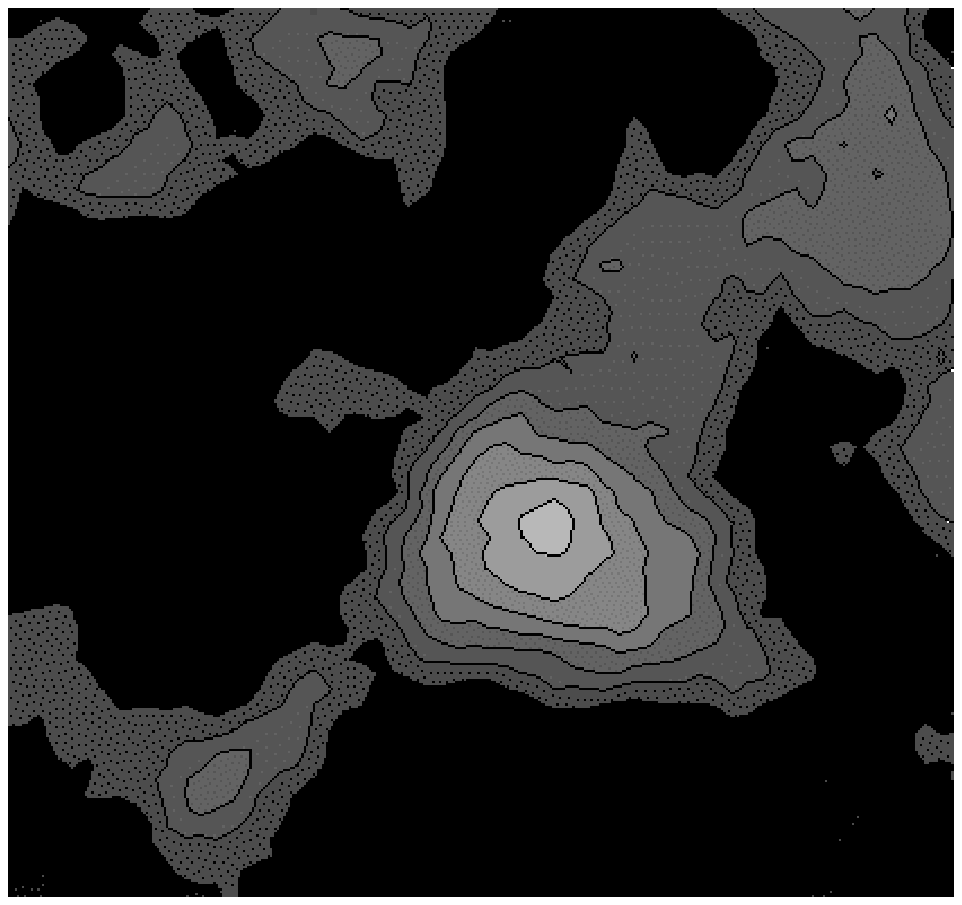}{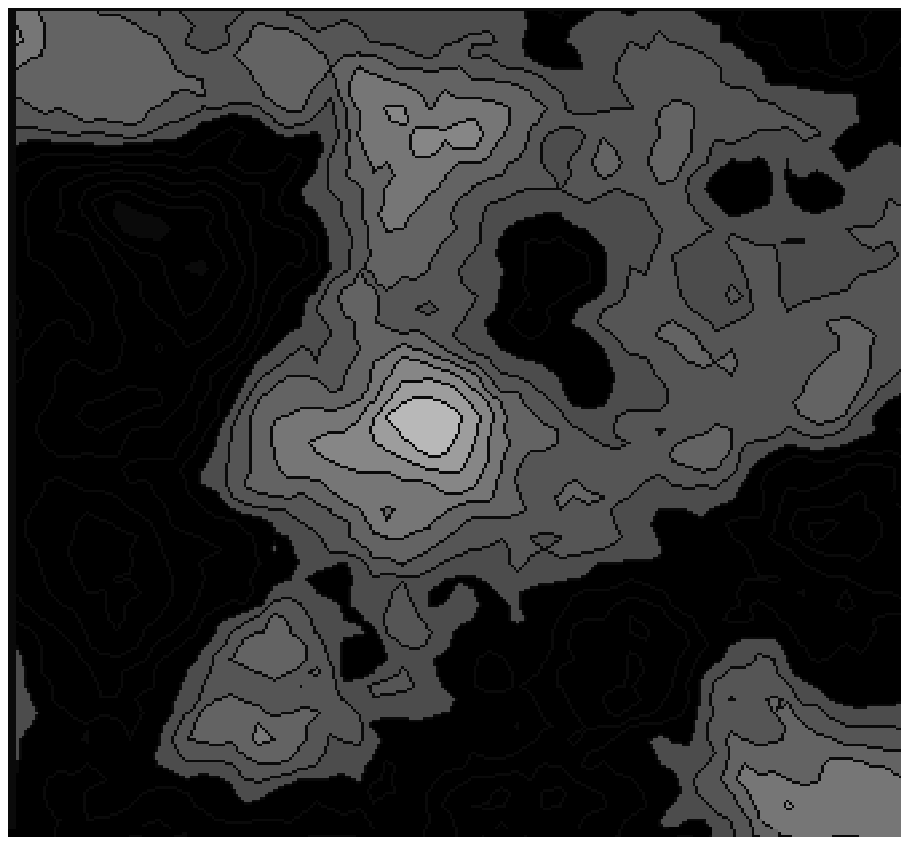} 
\caption{Projected mass maps of Abell 3266 and Abell 3667.  Contours are in signal to
noise with only positive contours greater than one sigma plotted. Both images are $44'\times 44'$.}
\end{figure}

\end{document}